# Enhancing Alzheimer's Disease Prediction: A Novel Approach to Leveraging GAN-Augmented Data for Improved CNN Model Accuracy

Akshay Sunkara[1] Rajiv Morthala[1], Anav Jain[1], Srinjoy Ghose[1], and Santosh Morthala[2]

[1]:Univeristy of North Texas

[2]:University of Texas at Dallas

Denton, United States of America

{AkshaySunkara, RajivMorthala, AnavJain, SrinjoyGhose}@my.unt.edu

Santosh.Morthala@gmail.com

*Abstract*— **Alzheimer's Disease (AD) is a neurodegenerative disease affecting millions of individuals across the globe. As the prevalence of this disease continues to rise, early diagnosis is crucial to improve clinical outcomes. Neural networks, specifically Convolutional Neural Networks (CNNs), are promising tools for diagnosing individuals with Alzheimer's. However, neural networks such as ANNs and CNNs typically yield lower validation accuracies when fed lower quantities of data. Hence, Generative Adversarial Networks (GANs) can be utilized to synthesize data to augment these existing MRI datasets, potentially yielding higher validation accuracies. In this study, we use this principle while examining a novel application of the SSMI metric in selecting high-quality synthetic data generated by our GAN to compare its accuracies with shuffled data generated by our GAN. We observed that incorporating GANs with an SSMI metric returned the highest accuracies when compared to a traditional dataset.**

## I. INTRODUCTION

Alzheimer's Disease (AD) is a devastating neurodegenerative disease that progressively impairs cognitive function, affecting millions of people around the world. About 6.5 million Americans aged 65 years and older have AD, a number expected to grow significantly as the population ages [1]. This trend underlies the need for early diagnosis and intervention to manage the diseases effectively and maintain better outcomes. As such, neural networks have become very promising for diagnosing (AD) using MRI scanning, even though they often fail to process the image data effectively [2].

Convolutional Neural Networks (CNNs) offer a promising solution in addressing image processing deficiencies. The convolution operation in CNNs plays a crucial role in reducing the complexity of image data while preserving essential features. This process involves applying filters to localized regions of an image, where each filter computes an aggregating function—such as the average or weighted sum—across its receptive field [3]. By extracting these localized features, CNNs capture critical patterns, enhancing the model's efficiency in tasks like image recognition and classification [4]. As these filters are applied across multiple layers, CNNs progressively learn hierarchical representations of the image, enabling the network to identify increasingly complex patterns and relationships. This layered approach allows CNNs to excel in a variety of image-based tasks, including the accurate classification and detection of key features.

CNNs can analyze brain imaging data to distinguish between healthy individuals and those with Alzheimer's by identifying crucial biomarkers. However, an insufficient amount of data can lead to a lower predictive accuracy for the CNN [5].

Generative Adversarial Networks (GANs) serve as a solution for lower predictive accuracy. GANs can be used to create synthetic data that closely represents real-world data, effectively augmenting existing datasets. [6] This synthetic data generation can be used in AD research in MRI models, by enlarging AD MRI datasets for the training of CNN models. This will mitigate data scarcity, further improving the performance of CNNs in the prediction of AD.

The main objective of our research is to augment existing datasets using GANs to create additional synthetic data. This study, we intend to evaluate whether GANs can positively augment our dataset volume and diversity, and lead to increased model accuracy [7].

## II. METHODOLOGY

### A. CNN Design and Architecture

Our group began our experiment by testing the accuracy of our CNN model using MRI (magnetic resonance imaging) from patients diagnosed with AD as well as healthy controls. The architecture begins with an input layer that uses three channels (R, G, B). The initial convolutional layer outputs 16 channels using a 3x3 kernel and a padding of 1, followed by a Rectified Linear Unit (ReLU) activation function and a max pooling layer with a 2x2 kernel and stride. The second convolutional layer outputs 32, utilizing a 3x3 kernel and identical activation and pooling layers. The output from this layer is flattened into a 1D feature vector, which feeds into a fully connected layer with 128 units, followed by a ReLU activation function and a dropout layer with a 0.5 dropout rate. The final output layer corresponds to the number of classes in the dataset. Cross-entropy loss is employed to minimize error, and the model is optimized using the Adam optimizer with a learning rate of 0.001 and a weight decay of 0.01. The model underwent training for 10 epochs, during which the training data continuously updated the model weights, and the training loss was monitored.

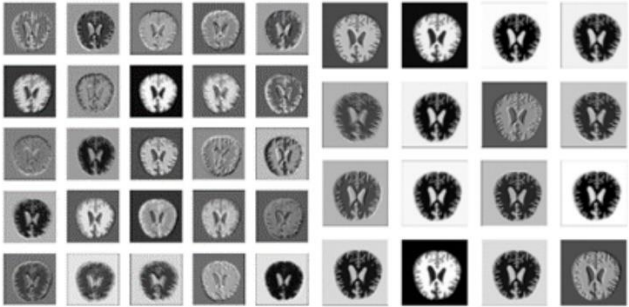

*Fig. 1: Feature Maps Derived from Convolution Layers*

*B. GAN-Enhanced Dataset*

We employed a GAN to generate additional MRI images to augment our dataset. The GAN model begins by importing MRI scans from patients with and without Alzheimer's disease. A directory is created for storing generated images, and a data transformation pipeline resizes images to 192x192 pixels, converts them to tensors, and normalizes them. The dataset is then loaded using ImageFolder. The DataLoader is created with a batch size of 32. The generator model accepts a 100-dimensional random vector and progressively upsamples it through multiple ConvTranspose2d layers, followed by ReLU activation and batch normalization, to produce a 3-channel output image. The discriminator is a CNN that classifies images as real or fake, using multiple Conv2d layers with LeakyReLU activations, dropout, and a fully connected layer with sigmoid activation to output probabilities.

During training, binary cross-entropy loss is used for both the generator and discriminator, optimized with Adam at varying learning rates. The discriminator is trained on real and generated images, while the generator is trained to produce images that can deceive the discriminator. Discriminator accuracy is closely monitored, and generated images are periodically saved throughout the training process. This approach aimed to enhance the diversity and volume of our training dataset, ultimately improving the predictive performance of our CNN model.

*C. SSIM Augmented Dataset*

The selection of the most accurately generated MRI scans from the GAN output was a major aspect of our study. We used the structural similarity index to evaluate differences between generated and reference images from the Alzheimer's disease dataset concerning luminance, contrast, and structure. We ranked the images according to their SSIM scores and chose the top 100 images from each class to keep only images proximal to the reference ones concerning these critical features. This approach would ensure a diversified variety and augment the volume of our training dataset to enhance the predictive performance of a CNN model.

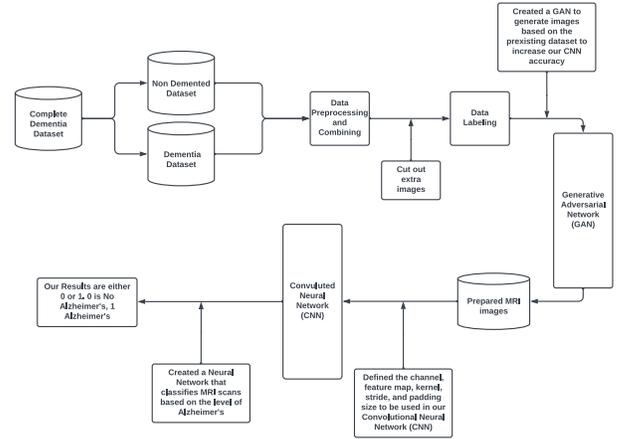

*Fig. 2: Flow Loop for Synthesizing Data & CNN*

## III. RESULTS

Subsequently, these selected images were utilized to train Convolutional Neural Networks (CNNs), examining the impact of synthetic data on model performance. We established a comparative framework consisting of three experimental conditions: training CNNs without synthetic data, with all generated data, and exclusively with the synthetic image with the highest SSIM score. Each condition was subjected to ten trials to robustly assess the influence of synthetic data, both with and without SSIM-based selection, on the validation accuracy of the CNNs.

The experimental results underscored the beneficial impact of synthetic images on CNN performance. During the trials, three distinct training setups were employed:

- No Synthetic Data
- Randomly Selected Synthetic Data
- SSMI Selected Synthetic Data

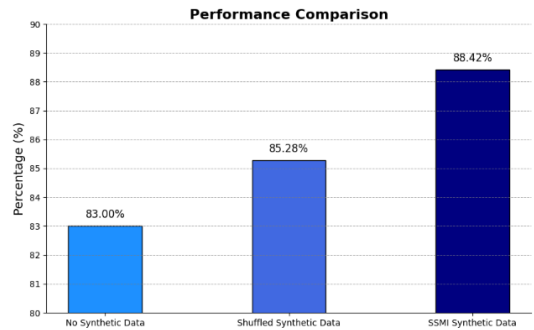

*Fig. 3: Performance Comparison Between Three Distinct Training Setups*

No Synthetic Data: In ten trials, each consisting of training the CNN for ten epochs on 200 images per class, the average accuracy observed was 83%, with peak accuracy nearly reaching 87.25% and the lowest at 60.75%.

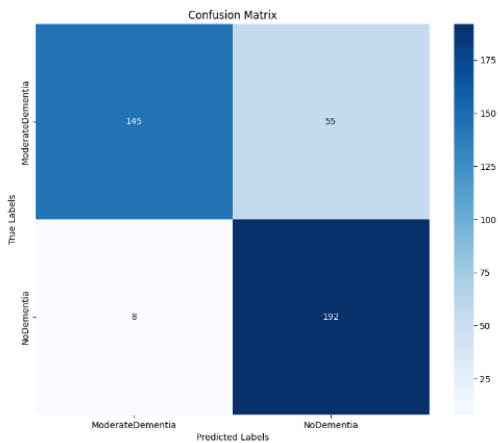

*Fig. 4: Confusion Matrix for CNN Without Synthetic Data*

Random Synthetic Inclusion: When the CNN was trained over the same number of epochs but with an augmented dataset comprising 250 images per class (including 50 randomly selected synthetic images), the results showed an average accuracy of 85.275%, with the highest accuracy of nearly 88.75% and the lowest 73.5%.

SSMI-Selected Synthetic Data: The most significant enhancement in model performance was noted when 50 of the highest SSMI-scored synthetic images were included in the training set of 250 images per class. This setup yielded an average accuracy of 88.425%, with the highest nearing 91.25% and the lowest accuracy at 84.5%.

These findings demonstrate that the highest accuracy was achieved by utilizing synthetic data selected based on SSMI scores for training CNNs, followed by the inclusion of randomly selected synthetic images. In terms of accuracy, the least effective approach involved training without synthetic data. This systematic use of the SSMI index to curate GAN-generated MRI scans significantly improved CNN's diagnostic accuracy.

## IV. CONCLUSION

*A. Future Works*

One significant enhancement in future iterations of our project could involve allowing our GAN model to run for extended durations. During the current experiment, our GAN was trained for 217 epochs. However, by increasing the number of epochs, we anticipate achieving even higher-quality MRI visualizations. Our observations indicated that the longer the GAN was trained, the more detailed and refined the resulting MRI visualizations became.

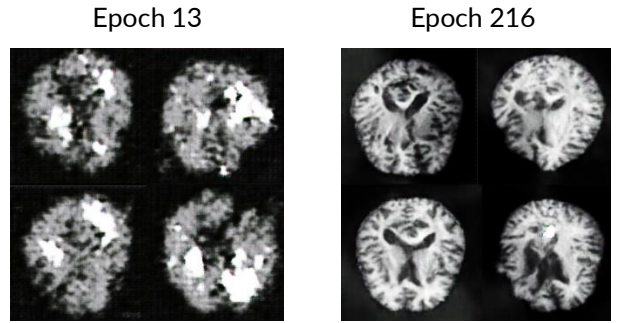

*Figure 5: GAN Generated MRI Scans*

Grad-CAM (Gradient-weighted Class Activation Mapping) is a technique used in deep learning. It helps to visualize where a convolutional neural network (CNN) is looking when making a prediction. Grad-CAM creates heatmaps that highlight important regions in an image [8].

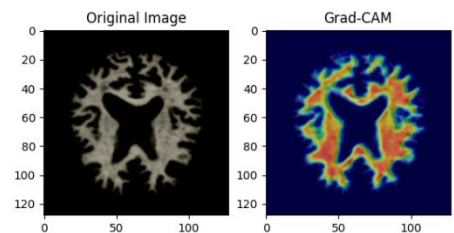

*Fig. 6: Grad CAM Image of MRI Scan from Dataset*

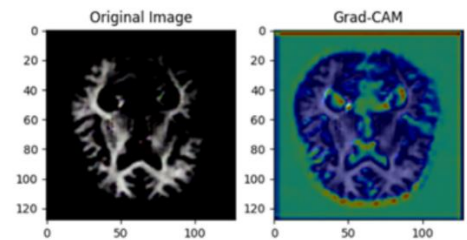

*Figure 7: Grad CAM Image of MRI Scan From GAN*

Based on these images, the GAN images have few spots that influence CNN's results. Training on higher epochs is probably required for the GAN to replicate the areas of importance seen in the original images.

Another key improvement we plan to explore is expanding the number of images selected by our SSMI. Currently, the SSMI selects 50 MRI images from patients with Alzheimer's disease and 50 images from those without the condition. We believe our CNN model could deliver more accurate and robust results by increasing the number of selected images.

Our findings indicate that using synthetic images in our dataset significantly enhances the accuracy of our convolutional neural networks (CNNs) validation. The technique of selecting synthetic images with the highest Structural Similarity Index (SSMI) produced the best validation accuracy across all experiments.

Moving forward, we also plan to perform parametric experiments on CNN performance to analyze the impact of various factors, such as the number of training epochs and the ratio of synthetic to non-synthetic images. We aim to refine the synthetic image generation process to closely mirror the characteristics of non-synthetic data, potentially implementing a feedback loop that uses SSMI metrics to guide further GAN training until the synthetic images achieve sufficient similarity. Throughout these experiments, we will carefully monitor for any signs of overfitting to ensure that enhancements in model accuracy are robust and reliable.

*B. Applications of Our Study*

Traditional methods of diagnosing Alzheimer's disease usually involve blood tests and lumbar punctures. Blood tests are potentially accurate diagnostic tools; however, many blood tests provide highly variable results, and their reliability in clinical settings is still being determined [9]. Lumbar punctures to extract cerebrospinal fluid are highly invasive procedures and can double the risk of bleeding in patients [10]. MRI imaging Our study proves that by incorporating CNN models along with GAN-augmented and SSMI-selected data, we can boost diagnostic accuracies to the same level as traditional diagnostic methods.

Another key finding from our study is the utilization of GAN-augmented data. By using GAN-synthesized MRI images, we improved the accuracy of our CNN model. Thus, future studies could incorporate GAN-augmented datasets to enhance their accuracy in diagnosing other diseases that can be diagnosed with images such as Parkinson's Disease.

REFERENCES

[1] "2024 alzheimer's disease facts and figures," Alzheimer's & dementia: the journal of the Alzheimer's Association, https://pubmed.ncbi.nlm.nih.gov/38689398/ (accessed Aug. 9, 2024).

[2] M. Salvi, U. R. Acharya, F. Molinari, and K. M. Meiburger, "The impact of pre- and post-image processing techniques on Deep learning frameworks: A comprehensive review for digital pathology image analysis," Computers in Biology and Medicine, vol. 128, p. 104129, Jan. 2021. doi:10.1016/j.compbiomed.2020.104129

[3] J. B. Bae et al., "Identification of alzheimer's disease using a convolutional neural network model based on T1-weighted magnetic resonance imaging," Scientific Reports, vol. 10, no. 1, Dec. 2020. doi:10.1038/s41598-020-79243-9

[4] M. Liu, D. Cheng, K. Wang, and Y. Wang, "Multi-modality cascaded convolutional neural networks for alzheimer's disease diagnosis," Neuroinformatics, vol. 16, no. 3–4, pp. 295–308, Mar. 2018. doi:10.1007/s12021-018-9370-4

[5] N. N. Arslan and D. Ozdemir, "Analysis of CNN models in classifying alzheimer's stages: Comparison and explainability examination of the proposed separable convolution-based neural network and Transfer Learning Models," Signal, Image and Video Processing, vol. 18, no. S1, pp. 447–461, Apr. 2024. doi:10.1007/s11760-024-03166-5

[6] C. Qu et al., "Diagnostic performance of Generative Adversarial Network-based deep learning methods for alzheimer's disease: A systematic review and meta-analysis," Frontiers in Aging Neuroscience, vol. 14, Apr. 2022. doi:10.3389/fnagi.2022.841696

[7] X. Zhou et al., "Enhancing magnetic resonance imaging-driven alzheimer's disease classification performance using generative Adversarial Learning," Alzheimer's Research & Therapy, vol. 13, no. 1, Mar. 2021. doi:10.1186/s13195-021-00797-5

[8] R. R. Selvaraju et al., "Grad-cam: Visual explanations from deep networks via gradient-based localization," International Journal of Computer Vision, vol. 128, no. 2, pp. 336–359, Oct. 2019. doi:10.1007/s11263-019-01228-7

[9] J. Hamilton, "New blood tests can help diagnose alzheimer's - but some aren't as accurate as others," NPR, https://www.npr.org/2024/01/10/1224041731/new-blood-tests-can-help-diagnose-alzheimers-but-some-arent-as-accurate-as-other (accessed Aug. 9, 2024).

[10] "Spinal taps carry higher risks for infants and elderly, study shows," ScienceDaily, https://www.sciencedaily.com/releases/2009/03/090318113559.htm#:~:text=The%20study%20also%20shows%20that,the%20middle%20of%20the%20lower (accessed Aug. 9, 2024).